\documentclass[epj,final]{svjour}
\usepackage[T1]{fontenc}
\usepackage[latin1]{inputenc}
\usepackage{graphics}

\makeatletter

\providecommand{\LyX}{L\kern-.1667em\lower.25em\hbox{Y}\kern-.125emX\@}
\newcommand{\noun}[1]{\textsc{#1}}

\makeatother

\begin{document}

\title{Passive Tracer Dynamics in 4 Point-Vortex Flow}

\author{A. Laforgia\inst{1} \and X. Leoncini\inst{2} \and
L. Kuznetsov\inst{3} \and G. M. Zaslavsky\inst{2,4}}

\institute {Dipartimento di Energetica
``S. Stecco'', Universit\`a degli Studi di Firenze, INFN and INFM, Via di
Santa Marta, 3 50139 Firenze, Italy \and Courant Institute of Mathematical
Sciences, New York University, 251 Mercer St., New York, NY 10012, USA \and
Lefschetz Center for Dynamical
Systems, Division of Applied Mathematics, Brown University, Providence, RI 02912,
USA \and Department of Physics, New York
University, 2-4 Washington Place, New York, NY 10003, USA }
\date{Received: date / Revised version: date}

\abstract{
The advection of passive tracers in a system of 4 identical point vortices is
studied when the motion of the vortices is chaotic. The phenomenon of vortex-pairing
has been observed and statistics of the pairing time is computed. The distribution
exhibits a power-law tail with exponent \( \sim 3.6 \) implying finite average
pairing time. This exponents is in agreement with its computed analytical estimate
of \( 3.5 \). Tracer motion is studied for a chosen initial condition of
the vortex system. Accessible phase space is investigated. The size of the cores
around the vortices is well approximated by the minimum inter-vortex distance
and stickiness to these cores is observed. We investigate the origin of stickiness
which we link to the phenomenon of vortex pairing and jumps of tracers between
cores. Motion within the core is considered and fluctuations are shown to scale
with tracer-vortex distance \( r \) as \( r^{6} \). No outward or inward diffusion
of tracers are observed. This investigation allows the separation of the accessible
phase space in four distinct regions, each with its own specific properties:
the region within the cores, the reunion of the periphery of all cores, the
region where vortex motion is restricted and finally the far-field region. We
speculate that the stickiness to the cores induced by vortex-pairings influences
the long-time behavior of tracers and their anomalous diffusion.
}

\maketitle

\section{Introduction}

The understanding of the motion of a passive tracer evolving in an unsteady
incompressible flow is fundamental due to its numerous applications in various
fields of research. They range from pure mathematical interest to geophysical
flows or chemical physics. The underlying problem is related to the Lagrangian
representation of the fluid evolution. This approach uncovered the phenomenon
of chaotic advection \cite{Aref84}-\cite{Crisanti92}, which refers to the
chaotic nature of Lagrangian trajectories in a non chaotic velocity field and
hence reflects a non-intuitive interplay between the Eulerian and Lagrangian
perspective. The ongoing interest in geophysical flows sustains interest in
two-dimensional models \cite{provenzale99}-\cite{Meleshko93}. In case of an
incompressible flow, the tracer's motion can be described by a non-autonomous
Hamiltonian. Another peculiarity of two-dimensional turbulent flows is the presence
of the inverse energy cascade, which results in the emergence of coherent vortices,
dominating the flow dynamics \cite{Benzi86}-\cite{Carnevale91}. In order to
tackle these problems, point vortices have been used with some success to approximate
the dynamics of finite-sized vortices \cite{Zabusky82}-\cite{VFuentes96},
as for instance in punctuated Hamiltonian models \cite{Carnevale91},\cite{Benzi92,Weiss99}.
Recent work shows that high-dimensional point vortex systems have both the features
of extremely high-dimensional as well as low-dimensional systems \cite{Weiss98};
moreover, the merging processes observed in decaying two-dimensional turbulence
have been shown to result from the interaction of a few number of close vortices
\cite{Zabusky96} and make the understanding of low dimensional vortex dynamics
an essential ingredient of the whole picture \cite{Aref99,Sire2000}.

Another aspect of the problem is related to transport properties, which for
various observations and models exhibit anomalous features. These anomalous
properties are linked to Levy-type processes and their generalizations \cite{Chernikov90}-\cite{Kovalyov2000}.
These properties are often related to the presence of coherent structures, which
can be identified from a Lagrangian perspective by the means of an analytic
criterion\cite{Haller2000}. In previous work, the advection in systems of three
point vortices evolving on the plane has been extensively investigated \cite{KZ98}-\cite{LKZpreprint}.
Three point-vortex systems on the plane have the advantage of being an integrable
system and often generate periodic flows (in co-rotating reference frame)\cite{Aref79}-\cite{Tavantzis88}.
This last property allows the use of Poincar\'e maps to investigate the phase
space of passive tracers whose motion belongs to the class of Hamiltonian systems
of \( 1-1/2 \) degree of freedom. A well-defined stochastic sea filled with
various islands of regular motion is observed and among these are special islands
also known as ``vortex cores'' surrounding each of the three vortices. Transport
in these systems is found to be anomalous, and the exponent characterizing the
second moment exhibit a universal value close to 3/2, in agreement with an analysis
involving fractional kinetics \cite{LKZpreprint}. In this light, the origin
of the anomalous properties and its multi-fractal nature is clearly linked to
the existence of islands within the stochastic sea and the phenomenon of stickiness
observed around them \cite{KZ2000,LKZpreprint}. Nevertheless the inherent periodic
nature of the motion of three vortices may be thought as artificial when considering
systems with more degree of freedom and therefore universal long time behavior
of transport properties may be considered as particular to these low-dimensional
periodic systems. 

The motion of \( N \) point vortices on the plane is generically chaotic for
\( N\ge 4 \) \cite{Novikov78}-\cite{Ziglin80}. However, a system of four
vortices remains a low dimensional system, but loses the periodic property of
three point-vortex systems and is in this sense a more realistic modelization
of observed low dimensional behavior.  The chaotization of the underlying flow
is expected to bring modifications to the transport properties studied in \cite{KZ98}-\cite{LKZpreprint};
the nature and relevance of the changes is although unknown. A precise study
of these questions is required and the answer consequently provides a way to
test the robustness of the results obtained with three vortices. 

In this paper we investigate numerically the motion of a passive tracer in the
field generated by four identical point vortices on the plane. This study follows
previous work found in Ref. \cite{Babiano94} and \cite{Boatto99}. In Ref.
\cite{Babiano94}, a first physical picture was given and the persistence of
vortex cores where tracers are trapped was clearly stated, while in Ref. \cite{Boatto99}
a rich extensive study is presented and emphasis is made on the quasi-regular
motion of tracers in the region far from the vortices. We note that in both
of these papers, the characterization of ``regular'' trajectories is defined
by the means of vanishing Lyapunov exponents. For instance, when a tracer is
trapped within cores, despite the core's chaotic motion, two nearby initial
conditions do not diverge exponentially. The goal of the present work is to
identify the different structures and different mechanisms which may influence
the transport properties of passive particles. Since parameter space is quite
large, the philosophy is essentially descriptive, and tracers motion are studied
for one arbitrary chosen initial condition of the vortex system. As the far
field region has been investigated \cite{Boatto99}, we focus on the motion
near or inside the cores, which we believe should be generic even for many vortex
systems. We will come to the issue of transport properties and finite-time Lyapunov
exponents in a forthcoming publication.  

In Sec. \ref{section2}, we describe briefly the motion of four vortices. We
present a Poincar\'e section of the vortex system, which provides a good test
to our numerical integration and allows to characterize easily the chaotic or
non chaotic nature of the motion. A new section capturing all equivalent physical
realizations of the flow is introduced. This section shows the existence of
non-uniformity in the phase space, which is linked to the permutation of vortices
and is related to the observation of vortex-pairing. Statistics on pairing times
are computed and exhibit power-law tails, implying finite average pairing time.
In Sec. \ref{section3} the motion of tracers is studied, \noun{}the presence
of cores is confirmed and stickiness to the vortex cores is observed. The influence
of vortex pairing is studied, which proves to be a good trapping (untrapping)
mechanism of tracers around the cores. Pairing of vortices allows a special
behavior of tracers jumping from one core to another core, which opens the possibility
in many vortex systems of special transport features resulting from jumps between
cores. The motion within the core is studied, dependence of fluctuations as
a function from the distance to the vortex are computed, and no typical diffusion
behavior is found.

\section{Vortex motion\label{section2}}

\subsection{Definitions}

The solution of the two-dimensional Euler equation, describing the dynamics
of a singular distribution of vorticity 
\begin{equation}
\label{vorticity}
\omega (z)=\sum _{\alpha =1}^{N}k_{\alpha }\delta \left( z-z_{\alpha }(t)\right) ,
\end{equation}
 where \( z \) locates a position in the complex plane, \( z_{\alpha }=x_{\alpha }+iy_{\alpha } \)
is the complex coordinate of the vortex \( \alpha  \), and \( k_{\alpha } \)
its strength, in an ideal incompressible two-dimensional fluid can be described
by a Hamiltonian system of \( N \) interacting particles (see for instance
\cite{Lamb45}), referred to as a system of \( N \) point vortices. The system's
evolution writes
\begin{equation}
\label{vortex.eq}
k_{\alpha }\dot{z}_{\alpha }=-2i\frac{\partial H}{\partial \bar{z}_{\alpha }}\: ,\hspace {10mm}\dot{\bar{z}}_{\alpha }=2i\frac{\partial H}{\partial (k_{\alpha }z_{\alpha })}\: ,(\alpha =1,\cdots ,N)\: ,
\end{equation}
 where the couple \( (k_{\alpha }z_{\alpha },\bar{z}_{\alpha }) \) are the
conjugate variables of the Hamiltonian \( H \). The nature of the interaction
depends on the geometry of the domain occupied by the fluid, for the case of
an unbounded plane, the resulting complex velocity field \( v(z,t) \) at position
\( z \) and time \( t \) given by the sum of the individual vortex contributions,
writes:
\begin{equation}
\label{velocity_{f}ield}
v(z,t)=\frac{1}{2\pi i}\sum _{\alpha =1}^{N}k_{\alpha }\frac{1}{\bar{z}-\bar{z}_{\alpha }(t)}.
\end{equation}
 and the Hamiltonian becomes
\begin{equation}
\label{Hamiltonianvortex}
H=-\frac{1}{2\pi }\sum _{\alpha >\beta }k_{\alpha }k_{\beta }\ln |z_{\alpha }-z_{\beta }|=-\frac{1}{4\pi }\ln \Lambda \: .
\end{equation}
 The translational and rotational invariance of \( H \), provides the motion
equations (\ref{vortex.eq}) three other conserved quantities besides the energy,
\begin{equation}
\label{constantofmotion1}
Q+iP=\sum ^{N}_{\alpha =1}k_{\alpha }z_{\alpha },\hspace {1.2cm}L^{2}=\sum _{\alpha =1}^{N}k_{\alpha }|z_{\alpha }|^{2}.
\end{equation}
Among the different constants of motion, there are three independent first integrals
in involution: \( H \), \( Q^{2}+P^{2} \) and \( L^{2} \), consequently the
motion of three vortices on the plane is always integrable and chaos arises
when \( N\ge 4 \).

\subsection{Canonical Transformations}

Due to the chaotic nature of 4-point vortex system, the understanding of vortex
motion necessitates a different approach than for integrable 3-vortex systems.
We follow Ref. \cite{ArefPomp82} and perform a canonical transformation of
the vortex coordinates. This transformation results in an effective system with
\( 2 \) degrees of freedom, providing a conceptually easier framework, best
suited for a detailed analysis of the motion of four identical vortices. For
instance, this transformation allows to perform well defined two-dimensional
Poincar\'e sections, from which the properties of the motion are analyzed. The
details of the transformation are not given here but the outline is the following,
using \( z_{\alpha } \) (\( \alpha =1,\cdots ,4 \)) as the complex coordinates
of the four vortices. 
\begin{equation}
\label{areftransform1}
(2J_{n})^{\frac{1}{2}}e^{i\theta _{n}}=N^{-\frac{1}{2}}\sum ^{N}_{\alpha =1}\exp (i(2\pi n/N)(\alpha -1))z_{\alpha }\: ,\hspace {10mm}n=0,\cdots ,N-1\: ,
\end{equation}
with new canonical variables
\begin{equation}
\label{newcanonicalvar}
\left\{ \begin{array}{ccc}
\phi _{1}=\frac{1}{2}(\theta _{1}-\theta _{3}), & \phi _{2}=\frac{1}{2}(\theta _{1}+\theta _{3})-\theta _{2}, & \phi _{3}=\theta _{2}\\
I_{1}=J_{1}-J_{3}, & I_{2}=J_{1}+J_{2}, & I_{3}=J_{1}+J_{2}+J_{3}\: .
\end{array}\right. 
\end{equation}
The resulting Hamiltonian is rather complicated but independent of \( \phi _{3} \),
meaning that \( I_{3}=\frac{1}{2}L^{2} \) is a constant of motion. We mention
that this transformation by preserving and making use of the constants of motion
is taking into account the continuous symmetries of the system. However, besides
these symmetries, the system is also invariant under the discrete group of permutations.
This last feature is particular to the situation of four identical vortices.
It has been shown that for a subgroup of these permutation, the effect of these
symmetries on the couple \( (I_{1},\phi _{1}) \) leads to simple linear transformations
( see \cite{ArefPomp82}, results are reproduced in Table \ref{tablepermut}),
but the effect of for instance the permutation \( (2,1,3,4) \) on the vortex
system (which can be thought of as a relabeling \( z'_{1}=z_{2} \), \( z'_{2}=z_{1} \),
\( z'_{3}=z_{3} \), \( z'_{4}=z_{4} \)), leads to no simple transformation
on \( (I_{1},\phi _{1}) \). The effect of these permutations on \( (I_{2},\phi _{2}) \)
does neither lead to simple transformations. We shall discuss some consequences
of these issues in the computation of Poincar\'e sections. 
\begin{table}[!h]
{\centering \begin{tabular}{|c|c|c|}
\hline 
permutation&
\( I_{1} \)&
\( \phi _{1} \)\\
\hline 
\hline 
\( (1,2,3,4) \)&
\( I_{1} \)&
\( \phi _{1} \)\\
\hline 
\( (4,1,2,3) \)&
\( I_{1} \)&
\( \phi _{1}-\pi /2 \)\\
\hline 
\( (2,3,4,1) \)&
\( I_{1} \)&
\( \phi _{1}+\pi /2 \)\\
\hline 
\( (2,1,4,3) \)&
\( -I_{1} \)&
\( -\phi _{1}+\pi /2 \)\\
\hline 
\( (3,4,1,2) \)&
\( I_{1} \)&
\( \phi _{1} \)\\
\hline 
\( (4,3,2,1) \)&
\( -I_{1} \)&
\( -\phi _{1}-\pi /2 \)\\
\hline 
\( (3,2,1,4) \)&
\( -I_{1} \)&
\( -\phi _{1} \)\\
\hline 
\( (1,4,3,2) \)&
\( -I_{1} \)&
\( -\phi _{1} \)\\
\hline 
\end{tabular}\par}

\caption{Effect of the subgroup \protect\( D_{4}\protect \) on the values of \protect\( (I_{1},\phi _{1})\protect \).\label{tablepermut}}
\end{table}

To summarize the results obtained in \cite{ArefPomp82}, the motion is in general
chaotic, except for some special initial conditions, for instance when the vortices
are forming a square the motion is periodic and the vortices rotate on a circle,
then symmetric deformation (\( z_{3}=-z_{1} \) and \( z_{4}=-z_{2} \)) of
the square lead to quasiperiodic motion (periodic motion in a given rotating
frame), see \cite{ArefPomp82} for the complete details.

\subsection{Poincar\'e sections}

As a prerequisite to our investigations on the passive tracer's motion, a basic
understanding of the vortex subsystem behavior is necessary. For this matter,
an arbitrary initial condition is chosen and a Poincar\'e section of the vortex
system is computed. The section is a tool which insures that for the considered
initial condition, the trajectory has the desired generic chaotic behavior.
To perform Poincar\'e sections of the system, we proceed as in \cite{ArefPomp82},
and use the set of canonical conjugated variables:
\begin{equation}
\label{conjr1p1r2p2}
\left\{ \begin{array}{cc}
R_{1}=\left( I_{1}+I_{3}\right) ^{\frac{1}{2}}\cos 2\phi _{1} & R_{2}=\left( I_{3}-I_{2}\right) ^{\frac{1}{2}}\sin 2\phi _{2}\\
P_{1}=\left( I_{1}+I_{3}\right) ^{\frac{1}{2}}\sin 2\phi _{1} & P_{2}=\left( I_{3}-I_{2}\right) ^{\frac{1}{2}}\cos 2\phi _{2}
\end{array}\right. \: ,
\end{equation}
 to compute the section \( R_{2}=0 \), \( \dot{R}_{2}<0 \). 

\begin{figure}[!h]
{\par\centering \resizebox*{8cm}{!}{\includegraphics{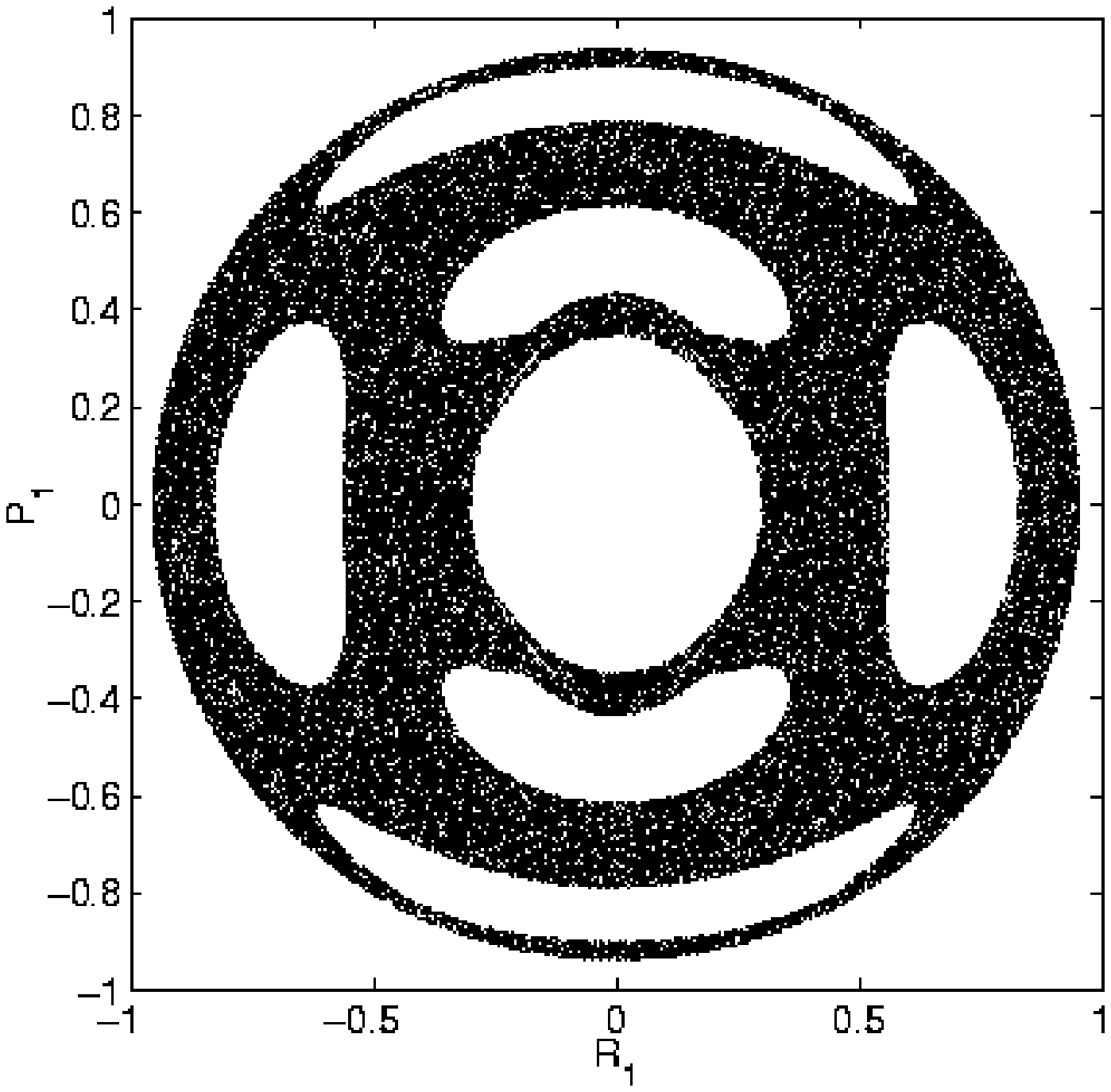}} \par}

\caption{Poincar\'e section of the four vortex system in the chaotic regime for a given
initial condition. The section is computed using the variables of the reduced
system (\ref{conjr1p1r2p2}). The conditions imposed are the following: \protect\( R_{2}=0\protect \),
\protect\( \dot{R}_{2}<0\protect \).\label{chaoticsec}}
\end{figure}

The result of this section for the arbitrary chosen initial condition is presented
in Fig. \ref{chaoticsec}. We notice that the section reproduces relatively
well the results obtained in \cite{PullSaff91}, where the section was computed
using a numerical integration of the reduced system. This has the advantage
of offering a natural test of the accuracy of our numerical integration, which
is made in our case in the original vortex complex variables using a fifth order
simplectic Gauss-Legendre scheme \cite{McLachlan92}. Previous analogous sections
can be seen in \cite{ArefPomp82} and \cite{PullSaff91} and are all in good
agreement with our result. The motion of the vortices is chaotic. This settles
the choice of the initial condition of the vortex system, which we will use
from now on.
\vspace{0.375cm}

\begin{figure}[!h]
{\par\centering \resizebox*{8cm}{!}{\includegraphics{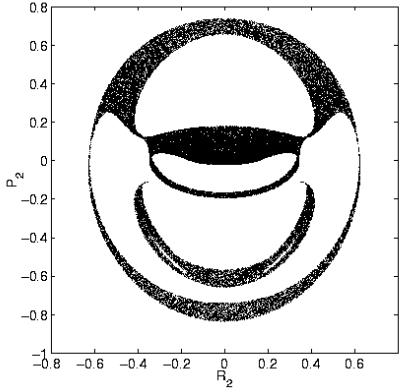}} \par}

{\par\centering \resizebox*{8cm}{!}{\includegraphics{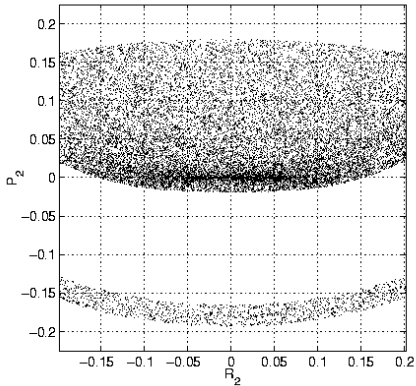}} \par}

\caption{Poincar\'e section of the four vortex system in the chaotic regime for a given
initial condition. The section is computed using the variables of the reduced
system (\ref{conjr1p1r2p2}). The conditions imposed are the following: \protect\( R_{1}=0\protect \),
\protect\( \dot{R}_{1}<0\protect \). This section has the advantage of containing
all equivalent physical configurations resulting from any permutation of the
vortices. Consequently stickiness is observed, while it was not in Fig. \ref{chaoticsec}.
The lower plot is a zoom of the upper one. \label{chaoticsec2}}
\end{figure}

We notice that Fig.\ref{chaoticsec} has symmetries which relate to the effect
of the permutations illustrated in Table \ref{tablepermut}. This somehow reflects,
as anticipated earlier on, that to equivalent physical configurations of the
vortex system correspond different points located in different regions of the
section. Moreover the fact that the new canonical variables are not invariant
under all permutation of the vortices implies that equivalent physical realizations
of the system have different location in the reduced phase space, and are likely
to not all belong to the section. From the advection point of view, equivalent
physical configurations generate identical flows, so the proposed section may
not be best suited for investigating advection and possible patterns. However
they prove very useful to characterize the type of motion (periodic, quasiperiodic,
chaotic). One possible way to circumvent this location problem follows from
noticing that the condition \( R_{1}=0 \) or \( P_{1}=0 \) remains unchanged
by the permutations listed in Table \ref{tablepermut}, and to capture all equivalent
systems, we can superimpose the two following sections \( R_{1}=0 \), \( \dot{R}_{1}<0 \),
and \( R'_{1}=0 \), \( \dot{R}'_{1}<0 \), where \( ' \) stands for the values
of \( R_{1} \) and \( \dot{R}_{1} \) obtained after the relabeling \( (2,1,3,4) \)
is performed; (note that the superimposition does not modify the obtained section
for the considered initial condition, but allows for more points and reduces
computation times). The corresponding section is plotted in Fig. \ref{chaoticsec2},
for the same initial conditions used in Fig. \ref{chaoticsec}. We notice some
differences in the symmetries and shapes, but also in the local density of points,
which suggests a stickiness phenomenon of the vortex system, and therefore a
possible intermittency of the flow. We note also that this phenomenon is not
related to stickiness close to a regular motion of the type \( z_{3}=-z_{1} \)
and \( z_{4}=-z_{2} \), as in this exact case we would have \( I_{3}=I_{2} \)
(see \cite{ArefPomp82}) and stickiness would have been observed in Fig. \ref{chaoticsec}.
In fact, stickiness is observed on the line \( P_{2}=0 \) around \( R_{2}=0 \),
so \( J_{2}\approx 0 \) and identically \( |z_{1}-z_{2}+z_{3}-z_{4}|\approx 0 \).
Since two vortices can not coincide a simple solution to this last equation
is requiring the vortices to be almost aligned. We obtain a similar situation
as the critical one observed for three identical vortices \cite{KZ98}, where
alignment of the vortices implied a permutation of two vortices. It then reasonable
to speculate that the stickiness observed in Fig. \ref{chaoticsec2} is induced
by permutations of two vortices.

\begin{figure}[!h]
{\par\centering \resizebox*{8cm}{!}{\includegraphics{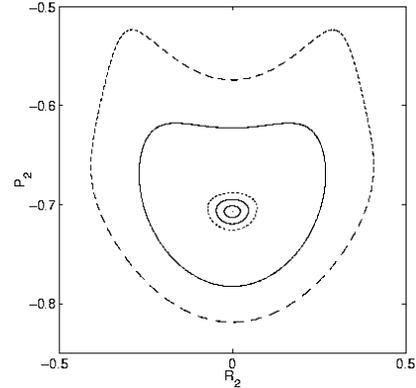}} \par}

\caption{Poincar\'e sections of the four vortex system in quasiperiodic regimes. The
different initial conditions correspond to continuous deformation from the square.
We notice the absence of chaos resulting in closed curves. The sections are
computed using the variables of the reduced system (\ref{conjr1p1r2p2}). The
conditions imposed are the following: \protect\( R_{1}=0\protect \), \protect\( \dot{R}_{1}<0\protect \).
\label{figsecvortalmostsquare}}
\end{figure}

The use of these sections cannot be more conclusive, and may be at this point
only a good indicator of some possible behavior of the vortex system. Indeed,
a simple glance at Eq. (\ref{conjr1p1r2p2}) shows that the conditions \( R_{1}=0 \)
or \( R_{2}=0 \) are degenerate in \( \phi _{1} \) respectively \( \phi _{2} \).
This implies that the plotted sections are in fact a super-imposition of different
sheets, which may be at the origin of the different domains observed in Fig.
\ref{chaoticsec2} and could also explain the difference in the center part
of the plots between the chaotic sections presented in \cite{PullSaff91} and
\cite{ArefPomp82}. To resolve the matter, a more detailed analysis is required,
namely the order of the degeneracy has to be established, as well as its effect
on the couples (\( R_{1},P_{1} \)), (\( R_{2},P_{2} \)). But even though interesting,
this is beyond the scope of this paper. The sections have clearly established
the desired chaotic nature of the vortex motion for our choice of initial condition.
Indeed, for comparison to the chaotic case, we computed in figures \ref{figsecvortalmostsquare}
sections for different configurations corresponding to quasiperiodic trajectories:
the difference in the nature of the trajectories is clear.

\subsection{Vortex pairing}

As mentioned, we use the initial condition corresponding to the chaotic trajectories
of the vortices (Fig. \ref{chaoticsec}). The Poincar\'e section illustrated
in Fig. \ref{chaoticsec2} suggests that the motion even though chaotic may
exhibit some intermittency and for a given length of time its behavior is similar
to an integrable system.  As mentioned earlier on, quasi-periodic motion of
four vortices can be observed for special initials conditions\cite{ArefPomp82}.
Essentially the initial conditions have to be symmetrical, and since all vortices
are identical the symmetry is preserved by the dynamics. This results in the
loss of degrees of freedom, and allows thus for an integrable motion of the
vortices. We may say that while the vortex system sticks to some domain of the
phase space, it stops exploring the whole accessible phase space but remains
on (sticks to) an object of lesser dimension than the total amount of degrees
of freedom. In this light, stickiness can then occur when two vortex are pairing.
Namely during the chaotic motion of the vortices, two vortices come close together
and form a pair. The pair is for a while acting like a two-vortex system perturbed
by the flow created by the two other vortices. While the pair is formed the
two vortices are bound, the systems looses one degree of freedom, and as a consequence
is sticking to some subdomain of the phase space. We may assume that this scenario
is more likely to appear than sticking to a symmetric configuration as in the
latter, the system needs to ``loose'' more degrees of freedom. 

\begin{figure}[!h]
{\par\centering \resizebox*{8cm}{!}{\includegraphics{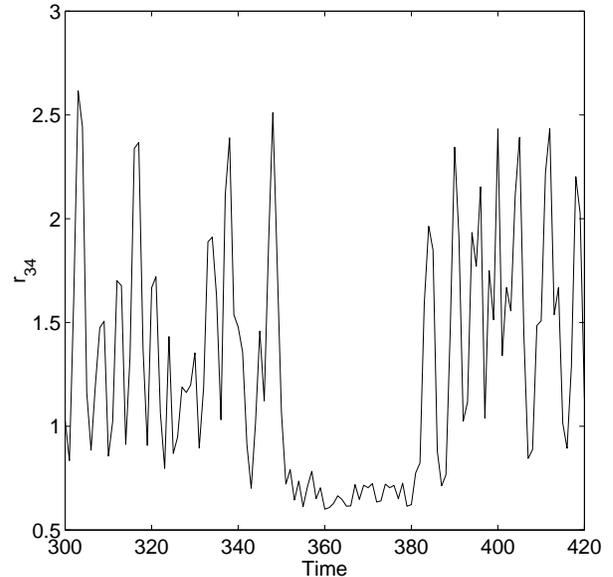}} \par}

\caption{Typical behavior of distance between vortices versus time. The distance \protect\( r_{34}\protect \)
is plotted. A pairing of the two vortices is identified for \protect\( 350<t<385\protect \).
We notice that while the pair is formed the two vortices remain close to each
other (\protect\( r_{34}<1\protect \)), and the fluctuations are greatly reduced.
This allows a simple diagnostic to numerically detect vortex-pairing \label{pairingillust}}
\end{figure}

To detect if vortex-pairing occurs, the inter-vortex distance is measured while
the evolution of the vortex is computed. The results are represented on Fig.
\ref{pairingillust} for a particular time frame, and as anticipated the pairing
of vortices is observed. For the illustrated time frame, two vortices are coming
close to each other around \( t=350 \), and remain at a distance \( d\approx 0.7 \)
from each other for a time \( \delta t\approx 30 \). We notice that while the
pairing occurs the inter-vortex distance remains small, and fluctuations are
greatly reduced. These features allows us to detect vortex-pairings in an easy
way. In regards to the previously computed Poincar\'e sections, it is important
to mention that while a pairing occurs the two vortices are rotating at a fast
pace around each other. This fast rotation is then likely to lead to ``quasi-permutations''
of the vortex system, and is probably the reason why we observe stickiness in
Fig. \ref{chaoticsec2}; since none is observed in Fig. \ref{chaoticsec}, we
confirm that stickiness to quasi-periodic motion resulting from symmetrical
initial conditions is less likely to happen than vortex-pairing.

\subsection{Statistics of the pairing time}

\begin{figure}[!h]
{\par\centering \resizebox*{8cm}{!}{\includegraphics{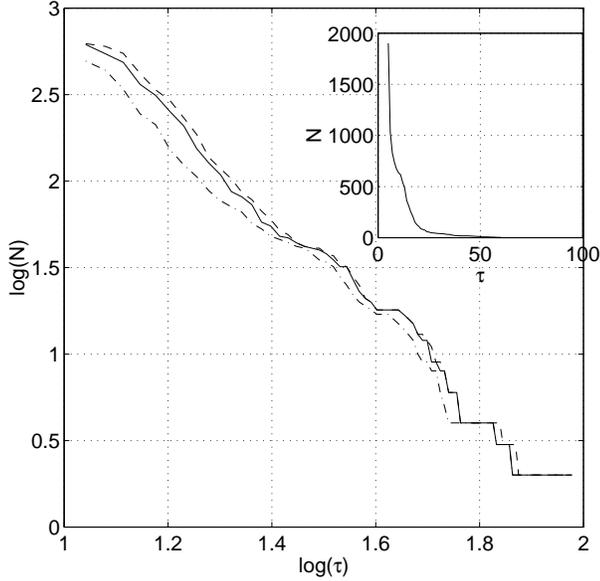}} \par}

\caption{Number of pairings \protect\( N\protect \) versus time length \protect\( \tau \protect \).
Only pairings lasting longer than \protect\( \tau =11\protect \) are considered.
The purpose of the different curves is to show the error-bar effects induced
by the chosen inter-vortex distance cutoffs \protect\( d=0.8,\: 0.9,\: 1\protect \)
which respectively correspond to the dashed-dotted, solid and dashed lines.
The tail of the curve shows a power law decay with coefficient \protect\( (\alpha -1)\approx -2.66\protect \).
The initial part of the curve arises error-bars. The run-time is \protect\( t=10^{5}\protect \).
\label{numofevents}}
\end{figure}

In previous work \cite{KZ2000,LKZpreprint}, it has been clearly shown that
stickiness by providing long coherent motion leads to anomalous transport properties,
and distributions with power law tails. As pairing can be considered a sticky
phenomenon and is very likely to influence the motion of passive tracers, it
is important to obtain some statistical data on pairing times. For this purpose,
a computation of vortex motion up to time \( t=10^{5} \) was made. The detection
of pairing events is obtained using Fig. \ref{pairingillust}: a pairing occurs
if for a given length of time two vortices stay close together. This definition
is rather vague and some arbitrary cutoff must be done. The arbitrary time length
chosen was \( \delta t=4 \), which does not affect the behavior of large pairing
time, and the distance from one vortex to another is such that \( r_{ij}=|z_{i}-z_{j}|\le d \).
To measure the influence of this last cutoff, we chose to use three different
values \( d=1 \), \( d=0.9 \) and \( d=0.8 \). We found enough pairings to
obtain a distribution of pairings that last longer than a time \( \tau  \),
meaning we computed the density
\begin{equation}
\label{probdef}
N(T>\tau )\sim \int ^{\infty }_{\tau }\rho (T)dT\: ,
\end{equation}
where \( \rho (T)dT \) is the probability of an event to last a time \( T \).
The results are shown in Fig. \ref{numofevents}, where we notice that some
rare events are lasting a relatively long time, and the different curves are
corresponding to different possible cutoff distance for \( d \), regarding
the detection of events. The analysis of the distribution's tail is done by
the Log-Log plot in Fig. \ref{numofevents}. A power-law decay of the tail of
the type \( \tau ^{-\alpha +1} \) with exponent \( \alpha \sim 3.66\pm 0.1 \)
is observed, which translates in non-zero probability of very long rare events:
long lasting pairing occurs. The behavior of the probability density of pairing
\( \rho (\tau ) \) lasting a time \( \tau  \), is obtained from Eq. (\ref{probdef})
\begin{equation}
\label{rhobeahvior}
\rho (\tau )=\frac{dN}{d\tau }\: ,
\end{equation}
which leads to a power law decay of the type \( \rho \sim \tau ^{-\alpha }\sim \tau ^{-3.7\pm 0.1} \).
This behavior translates in finite average pairing time, and second moment.
However since \( \alpha  \) is close to \( 4 \) and the accuracy obtained
for \( \alpha  \) is not perfect, we cannot be conclusive about the divergence
of the third moment. We insist that as expected, the long-lasting time-correlations
induced by a typical sticking behavior, results in a power-law decay of the
distribution's tail. We call \( \alpha  \) the pairing exponent and provide
its estimate in the following subsection

\subsection{Pairing exponent}

The main idea used to obtain an estimate of the value of the paring exponent
\( \alpha  \) follows the work presented in Ref.\cite{Zaslavsky2000} and \cite{LKZpreprint}.
The starting point is the occurrence of an island of stability resulting in
ballistic and accelerator modes. Such islands appear in the stochastic sea as
a result of a parabolic bifurcation \cite{Melnikov96} and correspond to the
so-called tangled islands \cite{romkedar99}. This is a fairly general statement
and it is hence reasonable to assume that the phenomenon of vortex-pairing results
from the rise of such islands in the stochastic sea, \emph{i.e} the formation
of virtual potential well for the rotational dynamics of a pair of vortices.
In this situation, we use the general form proposed in \cite{Melnikov96} and
write an effective Hamiltonian for the pair of vortices (see also \cite{Zaslavsky2000}and
\cite{ZEN97}):
\begin{equation}
\label{Heffecpair}
H_{eff}=c(\Delta P)^{2}-a\Delta Q_{1}-b\Delta Q_{2}-V_{3}(\Delta Q_{1},\Delta Q_{2})\:,
\end{equation}
where \( \Delta P \) is a generalized momentum (angular momentum) of the pair
and \( \Delta Q_{1},\Delta Q_{2} \) are the generalized coordinates of the
corresponding vortices. The interaction potential \( V_{3} \) is a third order
polynomial, as higher order terms in \( \Delta Q_{j} \) can be neglected for
the effective Hamiltonian {[}{]}; and \( a \), \( b \), \( c \) are constants.
Let us explain the expression (\ref{Heffecpair}) in more details.

Let us assume that the bifurcation corresponds to the appearance of an island
in the stochastic see at some phase space point \( \xi ^{*}=(P^{*}_{1},P^{*}_{2},Q^{*}_{1},Q^{*}_{2}) \).
After the bifurcation occurs, the island has a finite size and any trajectory
located within the island corresponds to periodic or quasiperiodic dynamics
characterized by its coordinates \( \xi =(P_{1},P_{2},Q_{1},Q_{2}) \). It is
then convenient to introduce the relative coordinates \( (\Delta P_{1},\Delta P_{2},\Delta Q_{1},\Delta Q_{2}) \)
by \( \Delta \xi =\xi -\xi ^{*} \). \noun{}Since the pair is rotating within
a plane, in the whole phase space, we can consider that angular momentum is
conserved, hence we only have one generalized momentum in Eq. (\ref{Heffecpair}),
while the linear and cubic terms are prescribed by the nature of the bifurcation.

The following steps are fairly formal (see also \cite{ZEN97} and \cite{LKZpreprint}).
Let us consider a trajectory which is close to the island's edge. A small perturbation
is then likely to allow the trajectory to ``escape'' from the island or its
vicinity and consequently to detroy the vortex pair. The phase volume of the
escaping trajectory writes:
\begin{equation}
\label{phasevolume}
\delta \Gamma =\delta P\delta Q_{1}\delta Q_{2}\:,
\end{equation}
where \( \delta P \) , \( \delta Q_{1} \), and \( \delta Q_{2} \) are the
values \( \Delta P \), \( \Delta Q_{1} \) and \( \Delta Q_{2} \) of the escaping
trajectory. Since the trajectory is close to the islands edge, we can estimate
from Eq.(\ref{Heffecpair})
\begin{equation}
\label{deltaPmax}
\delta P_{max}\sim \delta Q^{\frac{3}{2}}_{j},
\end{equation}
 where we have assumed \( V_{3}\sim Q^{3}_{j} \), \( (j=1,2) \). Using this
last expression (\ref{deltaPmax}), we obtain for (\ref{phasevolume})
\begin{equation}
\label{phasevolumebis}
\delta \Gamma =\delta Q^{\frac{3}{2}}_{j}\delta Q_{1}\delta Q_{2}\sim \delta Q^{\frac{7}{2}}\:,
\end{equation}
where we assumed \( \delta Q_{1}\sim \delta Q_{2}\sim \delta Q \). Due to the
periodic or quasiperiodic nature of the trajectories within the island, any
trajectory within its neighbourhoud will have a ballistic type behavior, hence
\( \delta Q\sim t \), \emph{i.e} 
\begin{equation}
\label{phasevolumeter}
\delta \Gamma \sim t^{\frac{7}{2}}\:.
\end{equation}
The probability density to escape the island vicinity after beeing in its neighbourhoud
for a time \( t \) (\emph{i.e} time-length of the pairing) within an interval
\( dt \) is 
\begin{equation}
\label{probdistbis}
\rho (t)\propto \frac{1}{\delta \Gamma (t)}\sim t^{-\frac{7}{2}}\:,
\end{equation}
 this results gives us directly the estimate of the exponent \( \alpha \approx 7/2 \),
which is very close to the observed value \( 3.7\pm 0.1 \).

Although this estimate is not rigorous, it can provide an insight on the origin
of different characteristic exponents of trapping time distributions.

\subsection{Minimum distance between two vortices}

To conclude on vortex motion, we measure the minimum distance between two vortices.
Indeed, as suggested in \cite{Babiano94}, the size of the cores surrounding
the vortices is related to the minimum distance of approach between two vortices
for a 3-vortex system \cite{LKZpreprint}. The minimum distance is numerically
measured and results are reported in Table \ref{Table2distmin}, which give
the value \( d_{min}\approx 0.6 \). An analytical estimation of the minimum
inter-vortex distance can be obtained (lower bound), by assuming that the minimum
occurs during a pairing, and that this minimum is small compared to the other
inter-vortex distances which we can assume to be all similar to a given distance
\( d_{av} \). Under these conditions the the constants of motions become 
\begin{equation}
\label{geomconditions}
K\equiv \left( \sum _{l=1}^{4}k_{l}\right) L^{2}-(Q^{2}+P^{2})=\sum _{l\neq m}^{4}k_{l}k_{m}|z_{i}-z_{j}|^{2}=d^{2}_{min}+5d^{2}_{av}\approx 5d^{2}_{av}\: ,
\end{equation}
and 
\begin{equation}
\label{Lambdabis}
\Lambda =\exp (-4\pi H)\approx d^{2}_{min}d^{10}_{av}\: .
\end{equation}
Using both equations (\ref{geomconditions}) and (\ref{Lambdabis}) we obtain
a simple estimation for the minimum inter-vortex distance
\begin{equation}
\label{dminexpression}
d_{min}=\sqrt{\left( \frac{5}{K}\right) ^{5}\Lambda }\: .
\end{equation}
We now use the expression (\ref{dminexpression}), with the values for the constant
of motions given by the initial positions of the vortices used for the simulations:
\( [(1.747,1.203) \) \( (-\sqrt2 /2,0) \) \( (\sqrt2 /2,0) \) \( (0,-1)] \).
This leads to \( d_{min}\approx 0.58 \), in very good agreement with the results
reported in Table \ref{Table2distmin}.

Having a rough picture of the underlying vortex-motion, we now focus on the
behavior of tracers.

\begin{table}[!h]
{\centering \begin{tabular}{|c|c|}
\hline 
Time of simulation&
Minimum distance\\
\hline 
\hline 
\( 10000 \)&
\( 0.6000 \)\\
\hline 
\( 20000 \)&
\( 0.5980 \)\\
\hline 
\( 30000 \)&
\( 0.5980 \)\\
\hline 
\( 50000 \)&
\( 0.5960 \)\\
\hline 
\( 100000 \)&
\( 0.5960 \)\\
\hline 
\end{tabular}\par}

\caption{Measured minimum distance between two vortices, with respect to simulation
time. We get roughly \protect\( \min (r_{ij})\sim 0.6\protect \), which translates
that the core's radius \protect\( r\protect \) is such that \protect\( r<0.3\protect \).
\label{Table2distmin}}
\end{table}

\section{Particle motion\label{section3}}

\subsection{Definitions}

The evolution of a tracer is given by the advection equation
\begin{equation}
\label{gen.adv}
\dot{z}=v(z,t)
\end{equation}
 where \( z(t) \) represent the position of the tracer at time \( t \), and
\( v(z,t) \) is the velocity field. For a point vortex system, the velocity
field is given by Eq. (\ref{velocity_{f}ield}), and equation (\ref{gen.adv})
can be rewritten in a Hamiltonian form: 
\begin{equation}
\dot{z}=-2i\frac{\partial \Psi }{\partial \bar{z}},\hspace {1.2cm}\dot{\bar{z}}=2i\frac{\partial \Psi }{\partial z}
\end{equation}
 where the stream function 
\begin{equation}
\label{stream}
\Psi (z,\bar{z},t)=-\frac{1}{2\pi }\sum _{\alpha =1}^{4}k_{\alpha }\ln |z-z_{\alpha }(t)|
\end{equation}
 acts as a Hamiltonian. The stream function depends on time through the vortex
coordinates \( z_{\alpha }(t) \), implying a non-autonomous system.

\subsection{Accessible phase space}

A first step in understanding the motion of passive tracers is to determine
their accessible domain in other words, where they move. Since the motion of
the vortices which are driving the flow is chaotic, we cannot use Poincar\'e
maps to visualize the phase space as in periodic 3-vortex systems. The possible
use of Poincar\'e recurrences as an analog to the period should also reveal
itself hazardous and numerically costly, as pairing was observed in the vortex
system, and stickiness is known to alter Poincar\'e recurrences statistics \cite{KZ2000,LKZpreprint}.
An alternative to phase space visualization was presented in \cite{Boatto99},
where the space of initial conditions was investigated by measuring for each
point its corresponding finite time Lyapunov exponent. This method was very
successful and allowed to visualize ``regular'' regions around vortices and
in the region far from the vortices, a very similar picture as the one observed
in three vortex systems. 

\begin{figure}[!h]
{\par\centering \resizebox*{8cm}{!}{\includegraphics{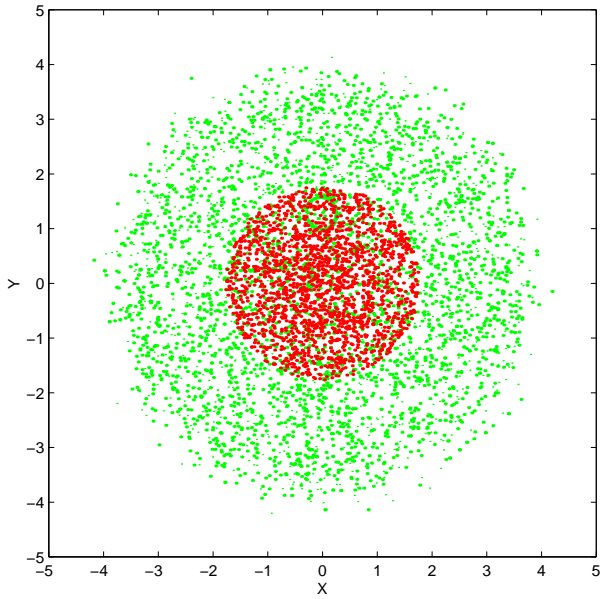}} \par}

\caption{Accessible phase space for the particles (gray), and the four vortices (dark).
The trajectories of the vortices and tracers are plotted on the plane. We notice
that the tracers have access to a broader domain in agreement with results found
in \cite{Boatto99}. The initial position of the tracers are taken in the region
of ``strong'' chaos.\label{Figaccessiblespace}}
\end{figure}

To confirm these results and get an idea of the accessible phase space; the
positions of vortices and tracers at different times are recorded and plot in
Fig. \ref{Figaccessiblespace}. This gives some first insights on the dynamics.
The motion of the vortices is confined within a circular region (`` the region
of strong chaos'' \cite{Boatto99}). In fact, the conservation of the angular
momentum \( L^{2} \) (see Eq. (\ref{constantofmotion1})) limits the vortex
motion on a hypersphere as all vortices are identical (infinite range of motion
can only be obtained with circulations of different signs). On the other hand
the motion of passive tracers evolves on a much wider range. This was previously
observed in Ref. \cite{Boatto99}, where it was shown that far from the region
where vortices are confined, tracers diffuse radially with a vanishing diffusion
rate. Contrary to three vortex systems, the chaotic nature of the vortex motions
destroys the barrier observed in quasiperiodic flows which limits the chaotic
sea to a finite region, and a diffusive regimes sets in.

We would like to emphasize that our choice of four identical vortices has been
based first on simplicity but also on the fact that the motion of the vortices
is then confined within a specific region of the plane centered around the center
of vorticity. If for instance we had changed the strength of one vortex to its
opposite value, sooner or later we would have observed the formation of a traveling
dipole \cite{Meleshko92}. Since we consider a flow on the entire plane, the
dipole will just travel towards infinity leaving behind a trivial system of
two vortices. The transport properties of such systems are then dramatically
modified, as the system becomes integrable as the dipole goes away, and just
transient chaos is expected.

\begin{figure}[!h]
{\par\centering \resizebox*{8cm}{!}{\includegraphics{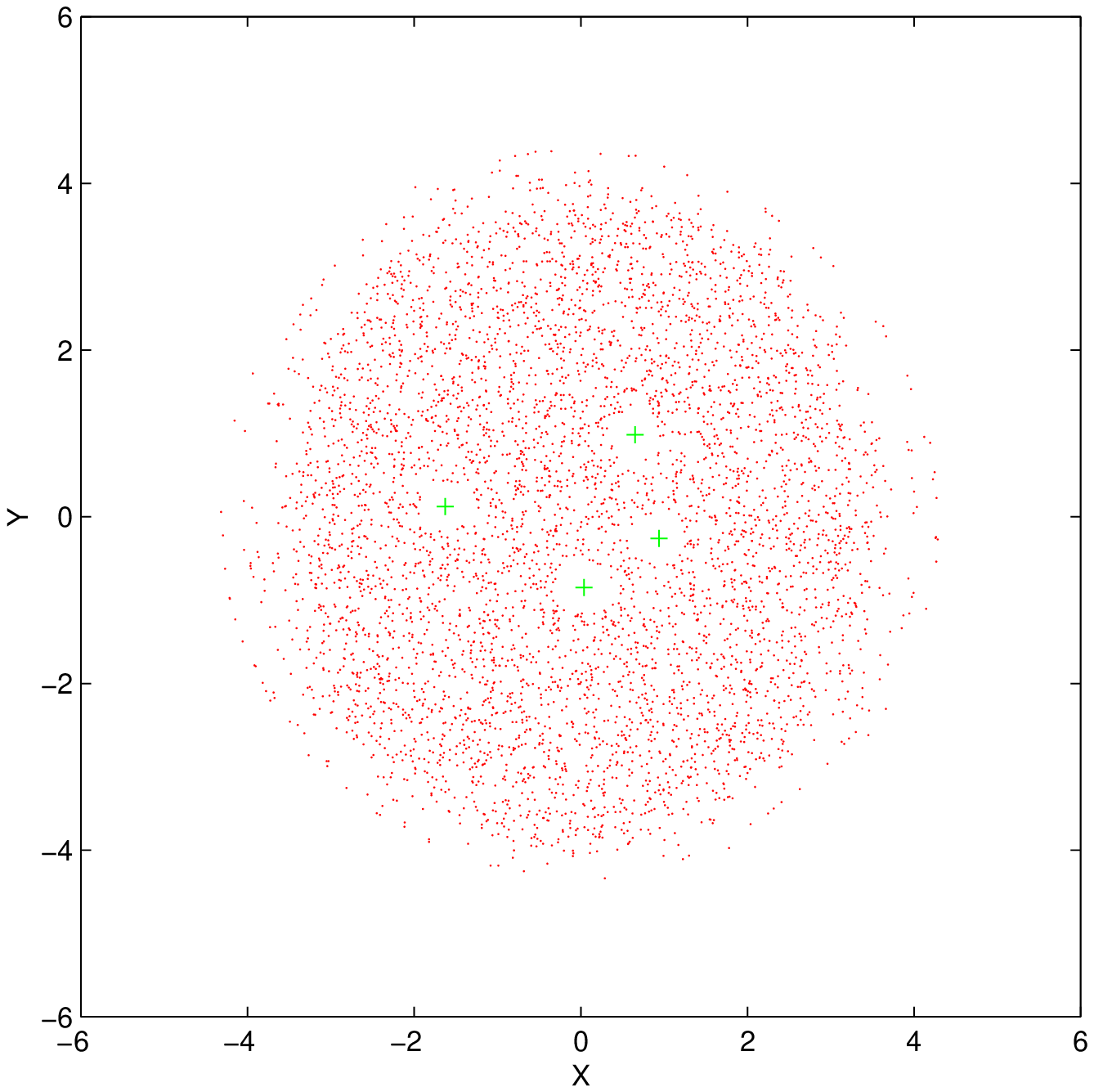}} \par}

{\par\centering \resizebox*{8cm}{!}{\includegraphics{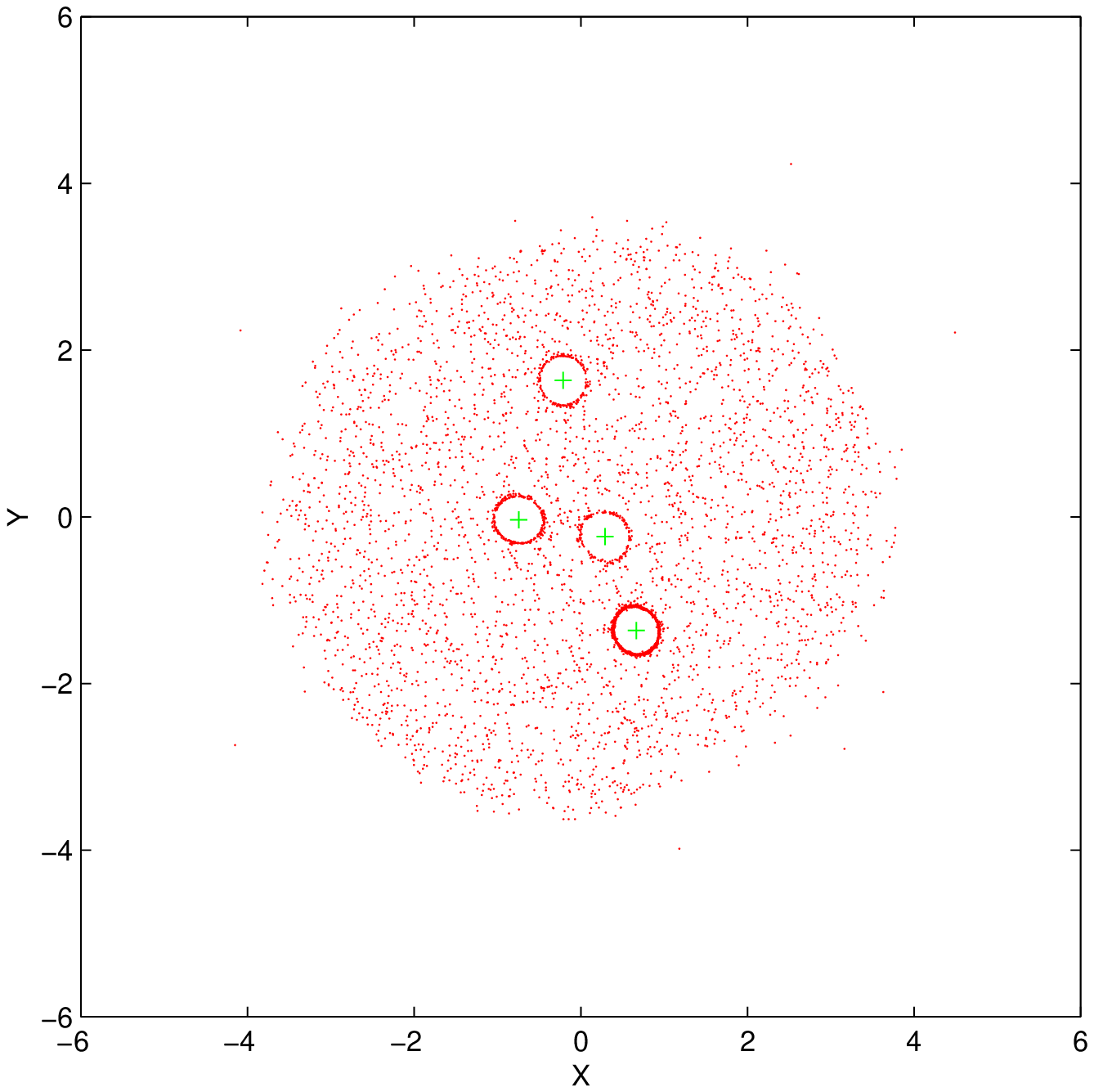}} \par}

\caption{Snapshots of a system with 5000 particles at two different times. The vortex
cores are showing, as expected. The core size is estimated around 0.3. On the
bottom plot, some tracers are sticking to the cores. We also notice the almost
regular motion described in \cite{Boatto99} for the far region. The initial
position of the tracers are taken in the stochastic sea.\label{sapshott550}}
\end{figure}

Further analysis of tracers motion is made using snapshots of the system at
different times, which was used in \cite{Babiano94} to visualize the cores
surrounding vortices. For this purpose \( 5000 \) passive tracers are initially
placed in the ``strong chaotic region'' where vortices evolve, and according
to Table \ref{Table2distmin}, at a distance larger than \( 0.3 \) from any
vortex. Results are presented in Fig. \ref{sapshott550}. As mentioned earlier
on for the case of quasi periodic flow with four vortices \cite{Boatto99} or
three \cite{KZ98,LKZpreprint}, circular-shaped islands of regular motion are
surrounding each vortex. This feature is preserved in the considered chaotic
flow. Although the motion of tracers within the core is probably not regular,
a barrier exists and prevents tracers to enter a a circular region around the
vortex (see Fig. \ref{sapshott550}), which we will refer to as vortex ``core''
from now on, regular motion or not. The presence of cores in this system  allows
to define de facto a Lagrangian size of a point vortex. It would be therefore
interesting to check if the Lagrangian definition of coherent structures boundary
given in Ref. \cite{Haller2000} gives the same results: namely are the cores
detected, and if so what is there size?

One snapshot illustrated Fig. \ref{sapshott550} shows also that at certain
times tracers can accumulate at the core's boundary. This property of the tracer's
motion is crucial and was not directly addressed in previous work. In fact,
this observation is very similar to the phenomenon of stickiness around vortex
cores observed in 3-vortex systems, and if the analogy retains, this form stickiness
should have a large influence on the transport properties of the flow, if not
essential when one focuses in the area of ``strong chaos'' \cite{KZ2000,LKZpreprint}.

\subsection{Stickiness around the vortex core}

The previous observation of both the existence of cores and the possible accumulation
of tracers around cores lead to further investigations of the cores surroundings.
First of all, one can wonder if the tracers can cross the barrier (if cores
are analogous to porous media), which could explain the observed accumulation.
Second, whether or not the barrier is crossed, it is important to know if a
typical (trapping) sticking time can be defined. 

A prerequisite to these questions is the estimation of the size of the core.
For this matter, we placed tracers at a given distance in the vicinity of one
vortex, computed its trajectory up to an arbitrary large time, and checked whether
or not the tracers remained in the vicinity of the chosen vortex. A simulation
was carried out for \( t=5.10^{4} \). One tracer placed at a distance \( r=0.25 \)
escapes around \( t\approx 3.10^{4} \), while it remains trapped for \( r=0.24 \).
Tracers, which are placed closer to the vortex than \( r=0.24 \), remain all
trapped, while tracers placed at a larger distance than \( r=0.25 \) escape
at smaller and smaller times. Therefore, given the considered initial conditions
and the time spans, we estimate the size of the cores \( r_{c} \) around \( 0.24 \).
And, as in the case of a three vortex system \cite{LKZpreprint}, we notice
that the measured value is is in good agreement with the expected upper limit
of \( r_{c} \) given by half the minimum inter-vortex distance (see Table \ref{Table2distmin}).
Moreover the fact that, as the distance from tracer to the vortex is increased,
the time of escape diminishes, seems a good indicator of a stickiness phenomenon.
However, the nature of the trapped tracer trajectory is unclear. A super-imposition
of the two trajectories computed for \( r=0.25 \) and \( r=0.24 \) reveals
a cross-over between the explored phase space, and therefore vanishing diffusion-like
processes with respect to \( r \) may be present. Namely, in the far region
vanishing-diffusion takes place. From a tracers point of view, the region were
vortices are confined becomes point like, and the flow resemble the one created
by one point vortex. Symmetrically, since inter-vortex distance are bound, as
the tracer is placed deeper inside the core, the flows becomes more one-point
vortex like; the difference with the far region being that the core has a chaotic
motion. 

Once the size of the core was determined, we tried to measure the escape time
\( T \) of a tracer as a function of the distance from the vortex \( r \),
expecting from the profile of \( T(r) \) to obtain a more accurate value of
\( r_{c} \) and information on sticking times. This attempt was unfruitful,
as the dependence of the escape time \( T \) on the initial position of the
tracer on the circle of radius \( \theta (t) \) (phase) \( r \) is very sensitive.
Another way to infer these properties was made by initializing a large number
of tracers uniformly distributed on a circle of radius \( r=0.24+\epsilon  \)
around one vortex, and obtain a distribution of escape times \( \rho _{e}(T,r) \).
This approach revealed itself also not convincing. Indeed, groups of particles
are escaping a given specific times, and the frequency of these escapes is not
high enough to allow us to obtain a smooth distribution \( \rho _{e}(T,r) \)
independent of the initial conditions of the vortex subsystem in a reasonable
amount of computation time.

\subsection{Tracer trapping (escaping) mechanism and core contamination}

\begin{figure}[!h]
{\par\centering \resizebox*{8cm}{!}{\includegraphics{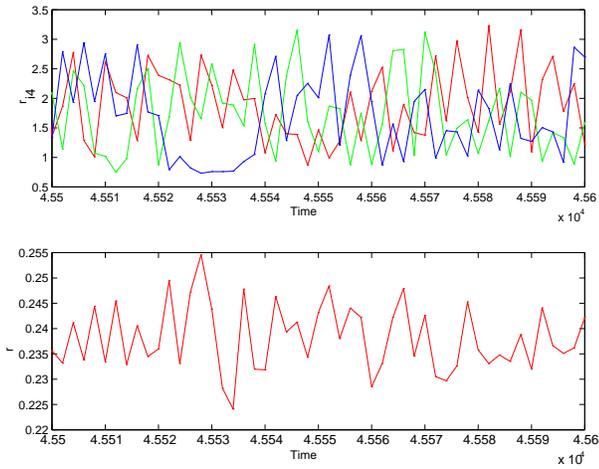}} \par}

\caption{Distance between one vortex and the others \protect\( r_{i4}\protect \) versus
time (upper figure). We notice the pairing of two vortices around \protect\( t=4.553\protect \)
\protect\( 10^{4}\protect \). It affects the behavior of the ``trapped''
tracer (bottom figure). The bottom figure shows the distance \protect\( r\protect \)
between one vortex and a passive tracer 'trapped' in the core versus time, while
the pairing occurs, fluctuations are amplified. Initial vortex positions is
\protect\( [(1.747,1.203)\protect \) \protect\( (-\sqrt2 /2,0)\protect \)
\protect\( (\sqrt2 /2,0)\protect \) \protect\( (0,-1)]\protect \). Initial
position of the tracer is \protect\( (0,-1.24)\protect \) (close to the fourth
vortex).\label{relativedistancevstime}}
\end{figure}

The observed sensitivity of escape times \( T \) on the phase of the tracer
and the strong discontinuity in \( T \) of preliminary measures of \( \rho _{e}(T,r) \)
suggests a non uniform behavior of the vortex system, with important consequences
on the vortex core surroundings. One potential candidate of this special behavior
is the pairing of two vortices observed earlier on, as the relative position
of the tracer to the second vortex and the sticky nature of vortex-pairing provide
a good origin to the difficulties arising in the previous attempts. In a first
step to check for this possible influence, a tracer was placed within the core.
Its relative distance from the attached vortex is measured concurrently with
the inter-vortex distances. The numerical data is collected in a time range
where a pairing occurs, and the tracer is trapped in one of the vortex forming
the pair. The time evolutions of the distances are presented in Fig. \ref{relativedistancevstime}.
This preliminary testings confirms that vortex-pairing has strong effects on
tracers motion, as while the pairing lasts, Fig. \ref{relativedistancevstime}
shows that fluctuations of the relative distance \( r(t) \) are increased.
We notice also that the increase of fluctuations goes without a noticeable jump
of the averaged distance from the vortex, which points to a strong dependence
on the phase of \( r(t) \) during a pairing.

\begin{figure}[!h]
{\par\centering \resizebox*{8cm}{!}{\includegraphics{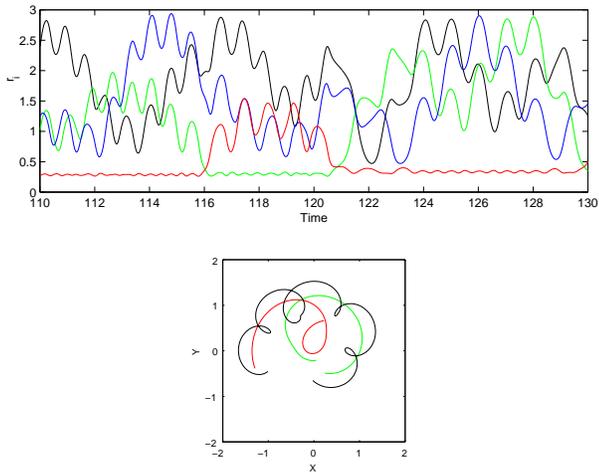}} \par}

\caption{The upper plot shows the distances \protect\( r_{i}\protect \) between one
advected particle and the four vortices. The tracer is initially placed close
to one vortex and sticks around the vortex for a certain time , then it jumps
and sticks onto another vortex . After a transition time it gets back on the
first vortex, we notice that the tracers has jumped to another orbit further
from the vortex. At the end of the plots the particle finally escapes from the
cores region. The bottom plot shows the trajectories in real space for \protect\( \delta t=4\protect \)
of time-interval ( from \protect\( t=114\protect \) to \protect\( 118\protect \)
). The black line refers to the trajectory of the particle, while the dark and
light gray lines to two the vortices involved in the jump. We note that while
the jump occurs there is a ``singularity'' on the trajectory of the advected
particle. Initial vortex position is \protect\( [(1.747,1.203)\protect \) \protect\( (-\sqrt2 /2,0)\protect \)
\protect\( (\sqrt2 /2,0)\protect \) \protect\( (0,-1)]\protect \). The initial
position of the tracer is \protect\( (0,-1.27)\protect \).\label{particlejump}}
\end{figure}

\begin{figure}[!h]
{\par\centering \resizebox*{8cm}{!}{\includegraphics{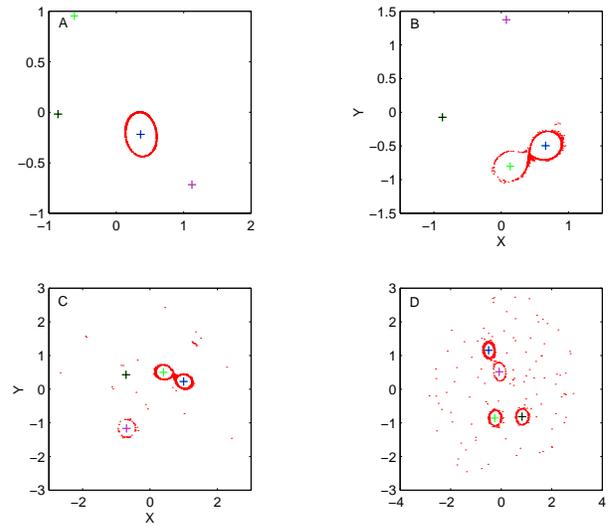}} \par}

\caption{Four consecutive snapshots for the four vortex of equal strengths and 1000
particles, corresponding to four consecutive pairing of the vortices. Even though
the particles are initially placed around one vortex, as pairings occur, some
of them jump from a vortex core to another and remain on them after the pairing.
While the vortex-pairing occurs some particles eventually escape from the cores.
We notice also that after four pairings all cores have been ``contaminated''
and are populated with tracers originating from the first core and while about
10\% of tracers have escaped from the region surrounding all four cores. Initial
vortex position is \protect\( [(1.747,1.203)\protect \) \protect\( (-\sqrt2 /2,0)\protect \)
\protect\( (\sqrt2 /2,0)\protect \) \protect\( (0,-1)]\protect \). Particles
are uniformly initialized on the circle of radius \protect\( r=0.27\protect \)
around the fourth vortex. \label{coreexchange}}
\end{figure}

\begin{figure}[!h]
{\par\centering \resizebox*{8cm}{!}{\includegraphics{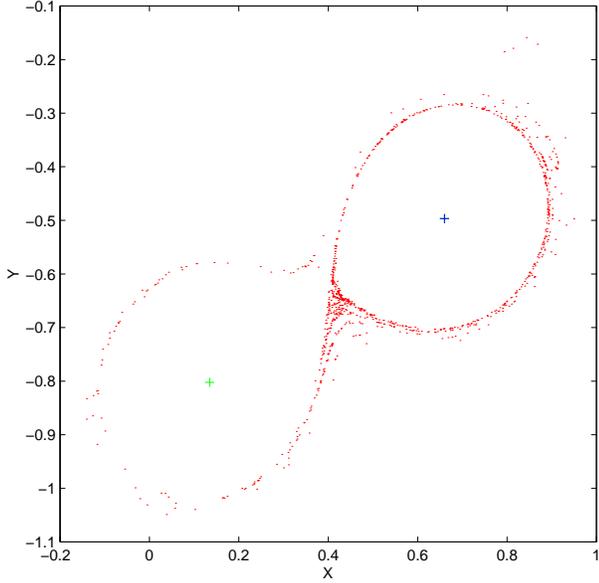}} \par}

\caption{Zoom of Fig. \ref{coreexchange} for the two vortex involved in the first pairing
(1000 particles). The vortex on the right is the closest to the initial position
of the particles. This kind of behavior is characteristic of pairing of two
vortices. During the pairing of the two vortices, the periphery of the cores
``merge'' and form a larger ``island'' where particles are trapped on and
may transfer from a vortex to another. Some particles may eventually escape
in the process. Initial vortex position is \protect\( [(1.747,1.203)\protect \)
\protect\( (-\sqrt2 /2,0)\protect \) \protect\( (\sqrt2 /2,0)\protect \) \protect\( (0,-1)]\protect \).
Particles are uniformly initialized on the circle of radius \protect\( r=0.27\protect \)
around the fourth vortex. \label{zoomcorexchange}}
\end{figure}

Continuing our step by step investigations, we initialized \( 1000 \) tracers
on a given circle of radius \( r=0.24+\epsilon  \) around one vortex and made
four different snapshots of the system, the first one before any pairing occurred,
and the last three in the middle of 3 consecutive pairings. The results are
shown on Fig. \ref{coreexchange}. This plot explain directly the observed discontinuity
\( \rho _{e}(T,r) \). We see that as pairing occurs the periphery of the two
cores merge, allowing tracers to jump from one core to the other. In this process
some tracers escape (and reversibly get trapped). When the pair is broken, each
cores keeps its share of exchanged tracers and only very few tracers escape
from the reformed cores. We also notice in Fig. \ref{coreexchange} that tracers
initially trapped on one core ``contaminate'' rapidly the other cores, and
we will refer to this phenomenon as ``core-contamination''. To better understand
this core-contamination process a zoom of the first exchange is presented in
Fig. \ref{zoomcorexchange}. While the two vortices are bound, the local topology
changes, and a peripheral merging of the cores occurs, allowing particle exchange
between cores. In this quasi 2-vortex system a quasi hyperbolic point in the
middle between the two vortices emerge, where the velocity field results from
the contribution of the two far vortices. As a result while the tracer finds
itself in this area, its motion is governed by the position of the two distant
vortices, allowing it to jump from one orbit to another and possibly one core
to another. A typical relative trajectory of a bound tracer is represented in
Fig. \ref{particlejump}. One notices that while the jump occurs the trajectory
is singular. 

These different events show the difficulty in defining properly a typical sticking
time from a distribution \( \rho (t,r) \), independent from the system's initial
conditions. First of all, while a tracer jumps from one core to another it continues
to stick, so the sticking area has to be considered globally on the four cores,
second even if particle do no jump from one core to the other, pairing allows
them to jump from one orbit to another, so the dependence on \( r \) of \( \rho  \)
is unclear. Finally, as the distribution of pairing times decays as a power-law,
rare events with long lasting pairings are not unprobable, implying a strong
lasting dependence on initial conditions for the computation of \( \rho (t,r) \).
For instance in Fig. \ref{particlejump}, after the last jump, the tracer finds
itself further from the center of the core, we also notice 3 aborted jumps at
\( t\approx 122,\: 123,\: 127 \), and finally around \( t=130 \), the tracer
escapes as a last vortex approaches. Consequently, trapped particles are subject
to escape from the cores as more pairing occurs, this illustrated in Fig. \ref{coreexchange},
where more and more tracers have escaped after each pairing.

We may speculate that this core-contamination phenomenon allows an unexpected
type of transport for the tracers in many vortex systems. Indeed, tracers which
find themselves in this peripheral zone, can jump from vortex to vortex, therefore
their transport properties should match the transport properties of the sub-vortex
system. Note also, that if all tracers are initially on one core like in Fig.
\ref{coreexchange} in a many vortex system the dispersion of tracers should
occur mainly on the different cores, implying non typical mixing properties
``targeted'' to the specific region of the phase space consisting of the reunion
of the periphery of all cores. This phenomenon, if persistent for more realistic
systems may have some interesting practical applications. Finally, we mention
that the sensitivity in these jumps from core to core in the relative position
on the core of the tracer \( \theta (t) \), must affect the computation of
finite time Lyapunov exponents in these regions, as their value will be influenced
by the chaoticity of \( \theta (t) \). It is therefore possible that non-zero
Lyapunov exponents are associated with trapped tracers, and we may expect a
typical value related to typical pairing frequency and pairing lifetime.

\subsection{Motion within the core}

\begin{figure}[!h]
{\par\centering \resizebox*{8cm}{!}{\includegraphics{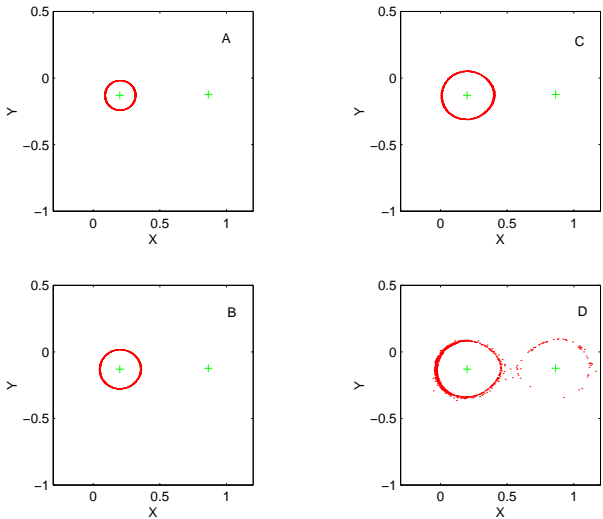}} \par}

\caption{Different snapshots with 1000 particles, initialized at different values of
the radius. The snapshots are taken while the first pairing occurs. An egg-shape
form of the cluster formed by the particles appears in plot C. Suggesting a
strong influence of pairing for \protect\( r>0.2\protect \). Initial vortex
position is \protect\( [(1.747,1.203)\protect \) \protect\( (-\sqrt2 /2,0)\protect \)
\protect\( (\sqrt2 /2,0)\protect \) \protect\( (0,-1)]\protect \). Radii for
the A,B,C,D plots are respectively \protect\( r=0.135\protect \), \protect\( r=0.18\protect \),
\protect\( r=0.225\protect \), \protect\( r=0.27\protect \).\label{insidecore}}
\end{figure}

\begin{figure}[!h]
{\par\centering \resizebox*{8cm}{!}{\includegraphics{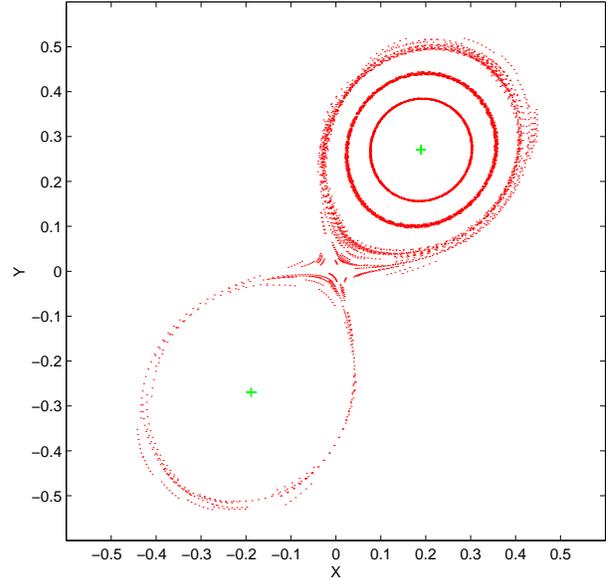}} \par}

\caption{Trajectories of 25 advected particles, for different initial conditions The
plot is made in the co-rotating frame with two vortices during a pairing lasting
\protect\( \Delta t=15\protect \), and particle positions recorded for constant
inter-vortex distance. Initial vortex position is \protect\( [(1.747,1.203)\protect \)
\protect\( (-\sqrt2 /2,0)\protect \) \protect\( (\sqrt2 /2,0)\protect \) \protect\( (0,-1)]\protect \).
Particles a uniformly distributed on circles of radii: \protect\( r=0.27\protect \),
\protect\( r=.20\protect \), \protect\( r=0.135\protect \). \label{poininsidecore}}
\end{figure}

We conclude the section by briefly investigating the motion of tracers within
the core. First the region within the core which is subject to a strong apparent
influence from pairing is localized. For this purpose, tracers are initialized
within the cores at different distances form the vortex, and a snapshot of the
system is made when two vortices are pairing in Fig. \ref{insidecore}. As expected,
trajectories look more and more circular as tracers are closer to the vortex,
and egg-shape deformation due to pairing appears for \( r>0.2 \). To confirm
this statement in Fig. \ref{poininsidecore} we plotted successive positions
of tracers during a pairing. The plot is made in a reference frame where the
two vortices do not rotate and centered on the local center of vorticity, the
position of the tracer is recorded for each time \( t_{i} \) such that the
distance between the two vortices \( r_{12}(t_{i}) \) is constant. This plot
is in spirit very similar to the Poincar\'e maps computed in \cite{KZ98}, and
gives a good insight on the local topology. However as pairing time is finite,
we typically obtain only \( \sim 10 \) iterations of the map, which limits
the resolution in Fig. \ref{poininsidecore}. 

\begin{figure}[!h]
{\par\centering \resizebox*{8cm}{!}{\includegraphics{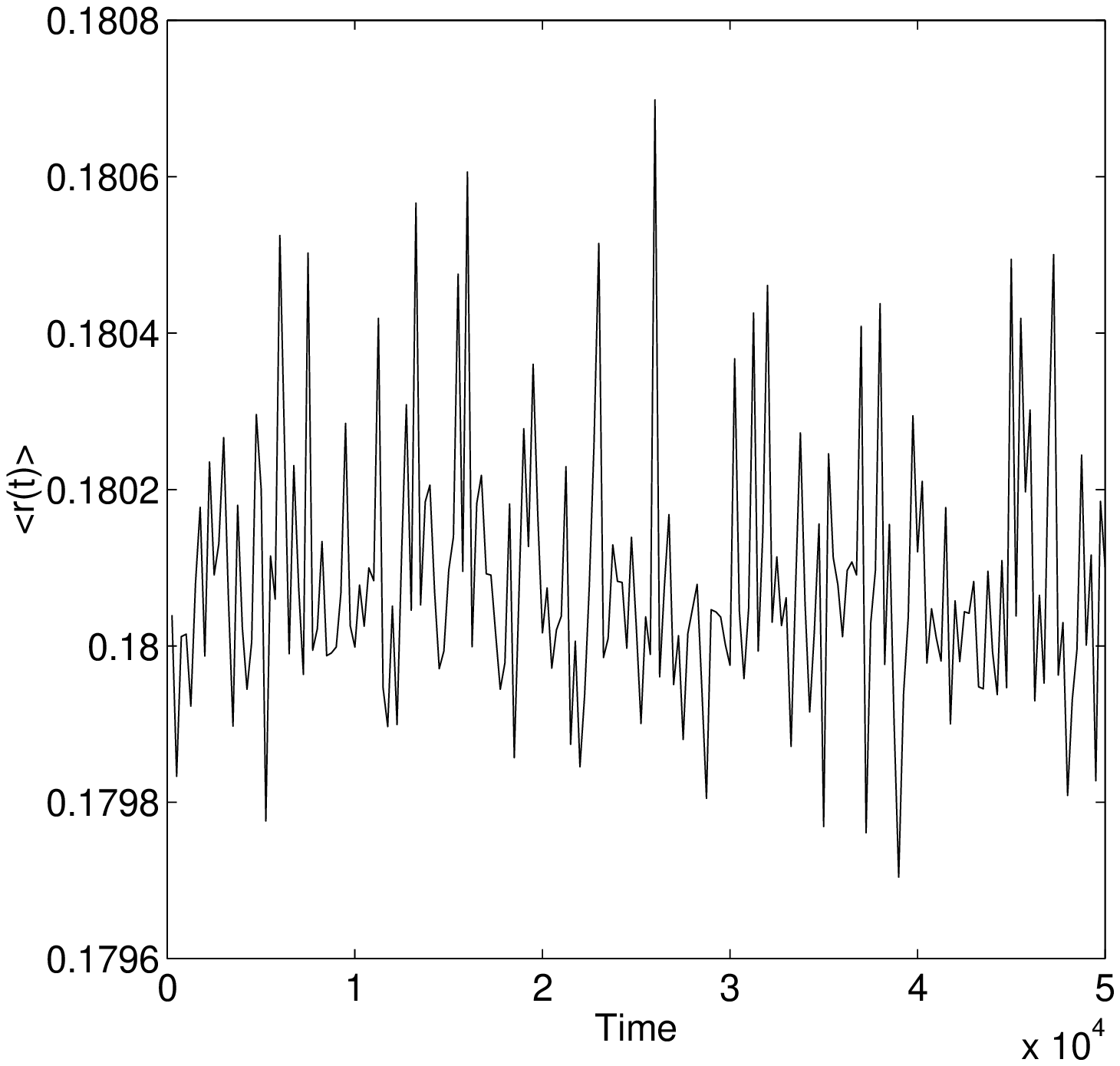}} \resizebox*{8cm}{!}{\includegraphics{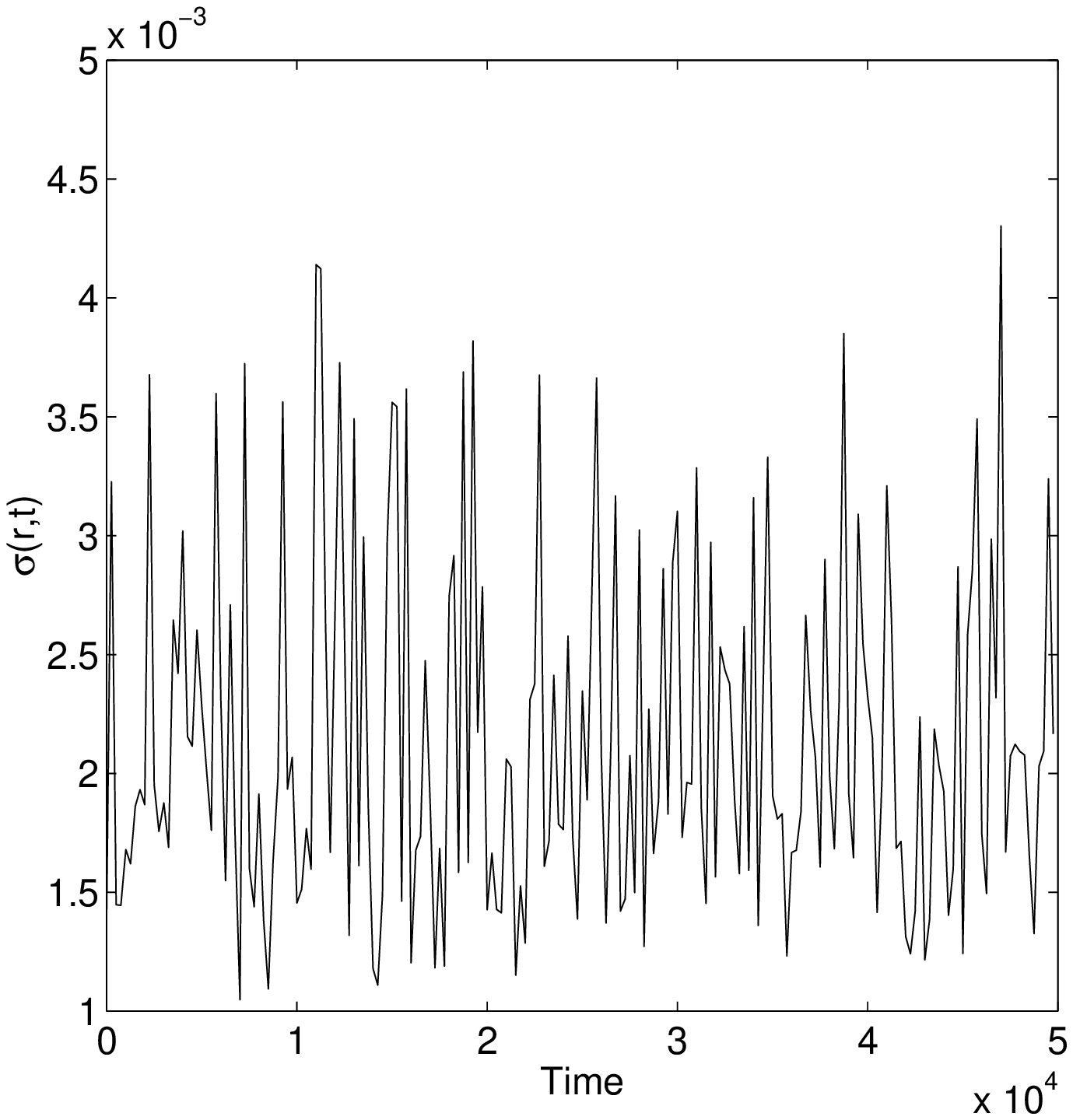}} \par}

\caption{On the left the mean value \protect\( \langle r(t)\rangle \protect \) as function
of time is plotted. The average \protect\( \langle r(t)\rangle \protect \)
can be considered constant in time. And on the right, the standard deviation
\protect\( \sigma (r,t)=\sqrt{\langle (r(t)-\langle r(t)\rangle )^{2}\rangle }\protect \)
as a function of time is plotted. We do not observe any quantitative diffusion
phenomenon, and conclude on the absence of diffusion for larger times span.
Data is computed using \protect\( 200\protect \) tracers. All tracers are uniformly
distributed at \protect\( t=0\protect \) on the circle of radius \protect\( r=0.18\protect \).
\label{fluctuationsinside}}
\end{figure}

Possible diffusion-like behavior of tracers need therefore be investigated deep
within the core. To detect this behavior, we initialized 200 tracers on a radius
\( r=0.18 \), and let the system evolve up to \( t=5.10^{4} \). The mean and
standard deviation \( \langle r(t)\rangle  \) and \( \sigma (r,t) \) are presented
in Fig. \ref{fluctuationsinside}. We notice that for the amount of particles
considered and the time-length of the simulation, no diffusive behavior is observed.
All particle remain trapped on the \( r=0.18 \) orbit. The absence of diffusion
is although not granted, but to be more conclusive we would need a larger amount
of tracers, and larger times. We are then confronted with numerical problems.
The divergence of the absolute speed as the vortex is approached, necessitate
increasingly smaller time steps, which lead to increasingly large effective
simulation times, and becomes an hindrance when one wants to compute statistics
over a large number of tracers whose trajectories are computed over large times.
Anyway for the time considered, the non observation of diffusion suggests that
the inside of the cores are regions of almost regular motion, if not regular,
and therefore cores are good trapping regions, which exchange little (if not
nothing) with the outside strong chaotic region.

\begin{figure}[!h]
{\par\centering \resizebox*{8cm}{!}{\includegraphics{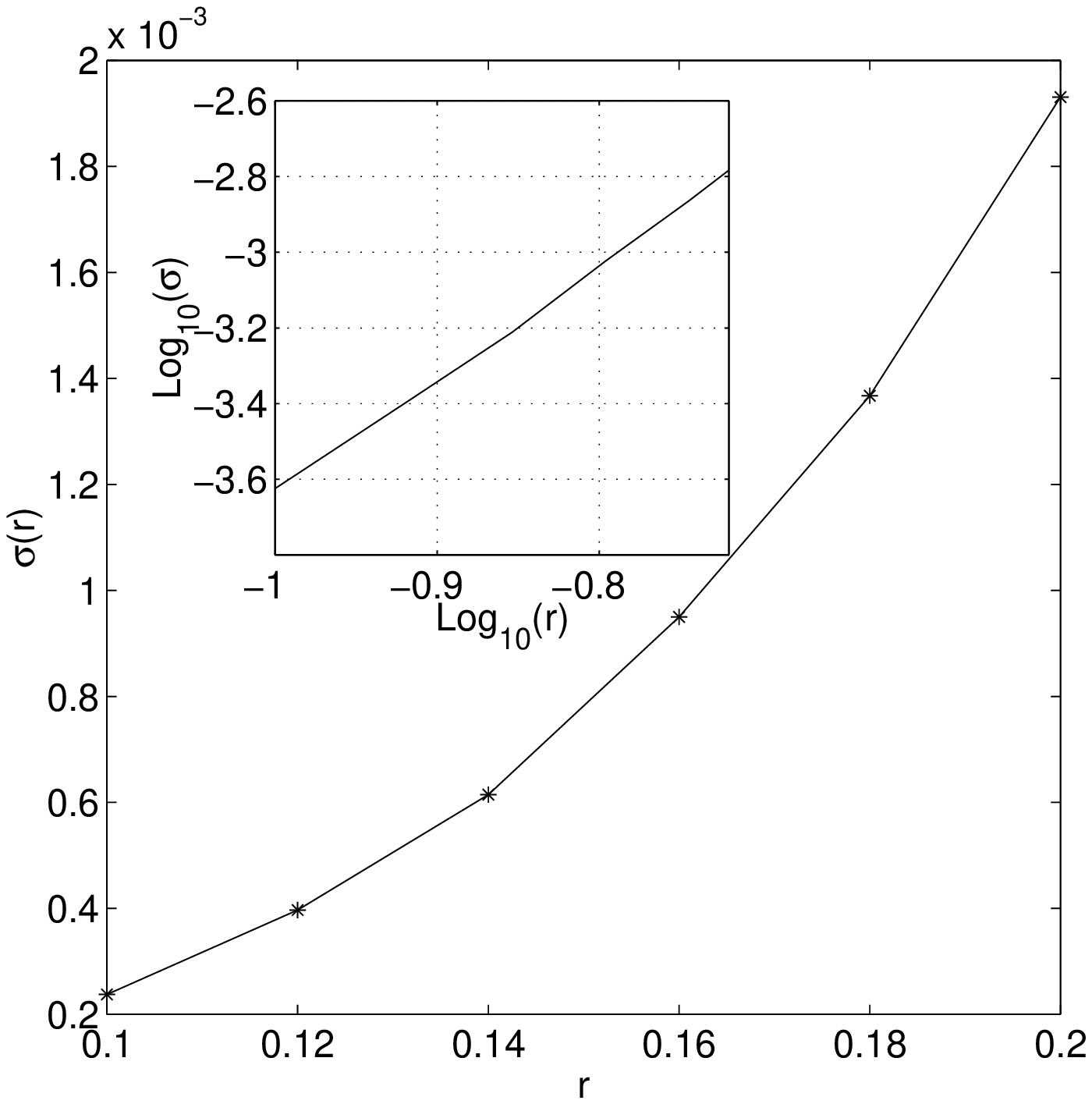}} \par}

{\par\centering \resizebox*{8cm}{!}{\includegraphics{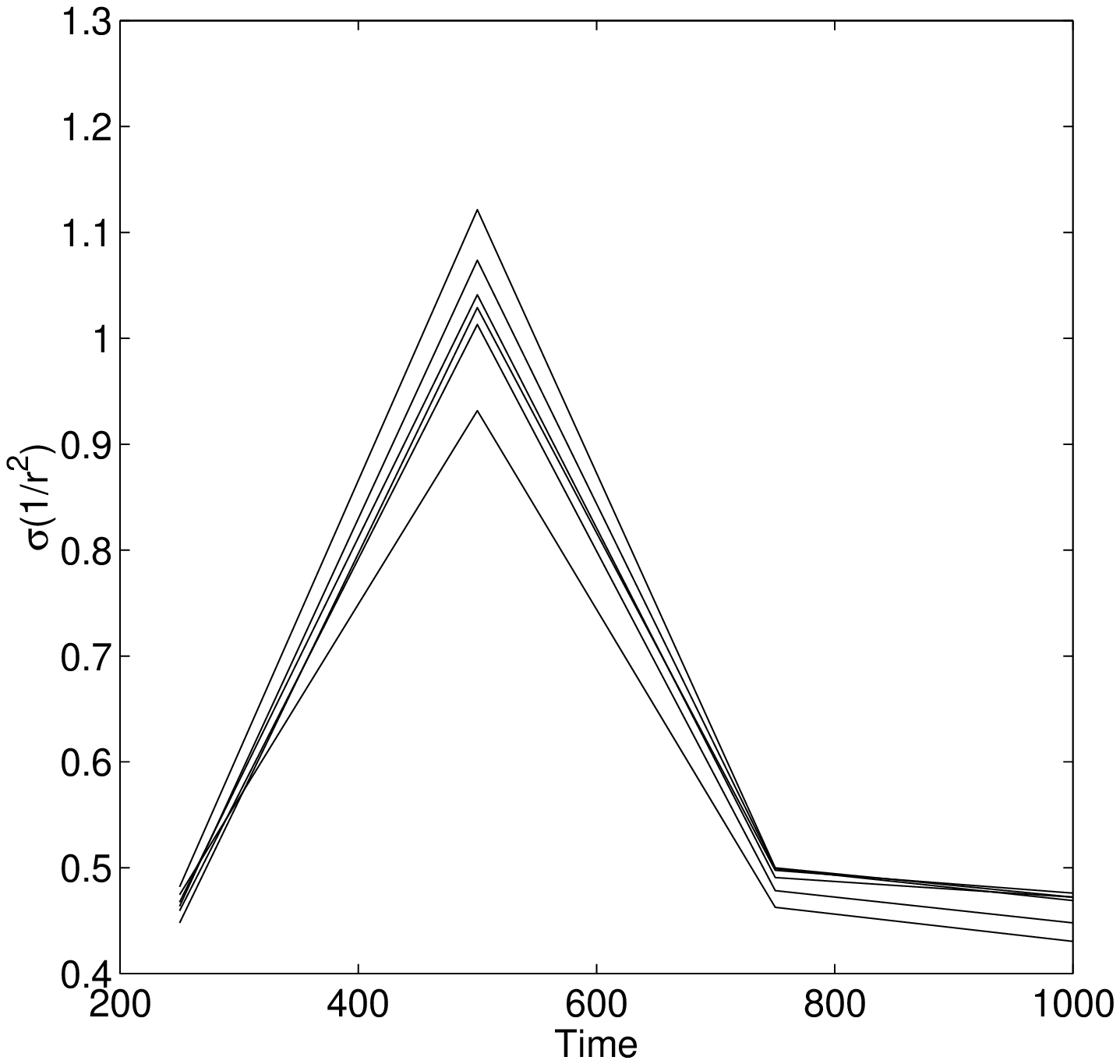}} \par}

\caption{On the upper plot the standard deviation \protect\( \sigma (r)=\sqrt{\langle (r-\langle r\rangle )^{2}\rangle }\protect \)
as a function of the distance from the center of the vortex \protect\( r\protect \)
is represented. The log-log plot suggests a behavior \protect\( \sigma \sim r^{2.93}\sim r^{3}\protect \).
On the bottom plot the standard deviation \protect\( \sigma (1/r^{2},t)\protect \)
is computed as a function of time for the different radii used in the upper
plot. All curves appear more or less to merge suggesting that \protect\( \sigma (1/r^{2},t)\protect \)
is only a function of time. Data is computed using \protect\( 500\protect \)
tracers. All tracers are uniformly distributed at \protect\( t=0\protect \)
on a circle of radius \protect\( r\protect \). \label{sigmavsr}}
\end{figure}

Finally the dependence of \( \sigma (r,t) \) as a function of \( r \) is investigated.
Results show that for up to \( r=0.2 \), \( \sigma (r)\sim r^{3} \) (see Fig.
\ref{sigmavsr}). This results is in fact very similar to the symmetry previously
discussed between what is observed in the far region \cite{Boatto99} and the
motion within the core. Namely this results translates into the fact that fluctuations
scale as \( \sim r^{6} \) where \( r \) is the small parameter, and in the
far field region it has been shown that the diffusion coefficient behaves as
\( D\sim 1/R^{6} \), where \( R \) is the distance from the center of vorticity
and \( 1/R \) is the small parameter. In fact we note that close to the vortex
we have \( \dot{\theta }\approx k/2\pi r^{2} \), therefore since \( \sigma (r,t)\sim r^{3} \),
we reasonably expect the standard deviation \( \sigma (k/2\pi r^{2},t)\approx \sigma (\dot{\theta },t) \)
to be only a function of \( t \). Results are shown on Fig. \ref{sigmavsr},
and we effectively notice that for all different radii, the fluctuations are
more or less concentrated on one curve. It is then sufficient to consider only
one orbit to study of the temporal behavior. Note also that for \( r=0.18 \),
we obtain \( \sigma (r)^{2}\sim 3.4\: 10^{-5} \), so for the observed time
if we had diffusion we should observe fluctuations of the order \( \sigma ^{2}(r,t)\sim \sigma ^{2}(r)t\approx 1 \)
for \( t=5.10^{4} \). In this light, since such growth of fluctuations was
not observed in Fig.\ref{fluctuationsinside}, we can exclude the possibility
of diffusive (or superdiffusive) behavior within the core. However, subdiffusive
behavior is still a possibility, as for instance \( (5.10^{4})^{1/8}\approx 3.87 \)
so the previous argument does not hold, and we need very long simulations to
clarify the temporal behavior of fluctuations.

\section{Conclusion}

In this paper we have investigated the motion of a passive tracer in a chaotic
flow generated by four identical point vortices. In the process of this study,
a particular attention has been made to the vortex subsystem. Since the vortices
are identical, the vortex system is subject to the additional discrete symmetry
resulting from invariance through the group of permutations. This lead to introduce
a new Poincar\'e section. This section has the advantage to exhibit the non
homogeneity of the phase space, resulting from a special behavior of the vortex
system corresponding to the phenomenon of vortex-pairing. Pairing results in
a temporary loss of a degree of freedom of the whole systems, it is therefore
associated with a stickiness phenomenon to an object of lesser dimension than
the accessible phase space, and is linked to local changes of density in the
Poincar\'e section illustrated in Fig. \ref{chaoticsec2}. Such behavior can
then be thought of as a chaos-chaos intermittent behavior. Vortex-pairing, by
involving only two vortices, is reminiscent of high-dimensional vortex systems
where low-dimensional vortex behavior was shown to be influential \cite{Weiss98}.
Further investigations have lead to compute pairing-time distributions. The
probability density exhibits a power-law tail typical the of stickiness behavior.
The power-law exponent is found to be around \( \alpha \approx 3.6 \), implying
finite typical (average) sticking time and quantitatively agree with its analytical
estimate \( \alpha \approx 7/2 \). We note that since most merging processes
in 2D turbulence occur while two same sign-vortices are pairing, the finiteness
of pairing time by the introduction of another specific time scale besides the
typical merging time, may play an important role. 

Passive tracers motion are analyzed for a given specific initial condition of
the vortices corresponding to a chaotic flow. The emphasis is made on qualitative
behavior as only one condition of the parameter space is explored and comparison
is made to previous work \cite{Babiano94,Boatto99} as well as results obtained
for quasiperiodic flows. The presence of cores surrounding the vortices is confirmed,
the sticking behavior of the tracers to the cores is illustrated, and an estimation
of the core size is measured, which is in good agreement with the estimation
proposed in Ref. \cite{Babiano94}, given by minimum inter-vortex distance.
The influence of vortex pairing on tracers dynamics is studied. This pairing
phenomenon proves to be a good trapping (untrapping) mechanism explaining the
observed stickiness of tracers around cores. Moreover, besides being responsible
for tracers trapping, vortex-pairing allows also tracers to jump between cores.
For the motion of the tracers located deep within the cores, we have shown that
for the time span studied, no radial diffusion-like behavior is observed. On
the other hand, the dependence of the fluctuations as a function of the distance
from the vortex is measured and these are shown to scale as \( r^{6} \), which
gives, in regards to the small parameter, the same order as the scaling previously
computed in the far field region in Ref. \cite{Boatto99}. 

To conclude, the motion of a passive tracer in the chaotic flow generated by
four point vortices has four different typical behaviors in four distinct regions
of the phase space. In the far field region, its motion is almost regular, with
some irregular jumps from one orbit to the other. Its motion is chaotic in the
region of ``strong chaos'', corresponding to the area where the vortex motion
is restricted (see \cite{Boatto99}). In the periphery of the cores, tracers
can stick to one core and travel with one specific vortex, as well as travel
through the phase space by jumping from one core to another. Finally deep inside
the vortex core, the tracers remain trapped on a specific orbit for the observed
time length of our simulations, and travel with the corresponding vortex. Transport
properties are subject to the influence of regions to which a tracer has access,
which counts all regions besides the inner of the cores. The existence of stickiness
on vortex cores and regular motion in the far field region allows us, by analogy
with three vortex systems \cite{KZ2000,LKZpreprint}, to speculate that long
time behavior of tracers is governed by these two regions. Each of them dominating
one part of the moments, namely the low moments for the far field region, and
the high moments for the periphery of cores. Moreover, the possibility for sticking
tracers to jump from core to core, leads to anticipate on the existence of special
transport features in many vortex systems. It is indeed likely that this feature
allows to transport highly concentrated regions of passive tracers over large
distances with controlled dilution and at fast pace. Namely, in many vortex
systems, contrary to tracers located within one core which are transported by
one and only vortex and hence may be trapped for relatively long times in some
slow moving cluster of vortices. Those located on the core's periphery should
quickly spread on all other cores periphery and therefore populate the whole
space accessible to the vortices; but by remaining on the cores, the spreading
should occur with relatively little dilution compared to the region of strong
chaos.

\section*{Acknowledgments}

We would like to thank the anonymous referee for his pertinent questions and
suggestions. One author, A. Laforgia would like to thank S. Ruffo for his constant
support and useful discussions. This work was supported by the US Department
of Navy, Grant No. N00014-96-1-0055, and the US Department of Energy, Grant
No. DE-FG02-92ER54184. This research was supported in part by NSF cooperative
agreement ACI-9619020 through computing resources provided by the National Partnership
for Advanced Computational Infrastructure at the San Diego Supercomputer Center.

\newpage

\newpage

\newpage

\newpage

\newpage

\newpage

\newpage

\newpage

\newpage

\newpage

\newpage

\newpage

\newpage

\newpage

\newpage

\end{document}